\documentclass[11pt,a4paper]{article}

\usepackage{graphicx}
\usepackage{amssymb,amsfonts,amsmath,mathtools}
\usepackage[font=small,labelfont=bf,labelsep=period,justification=raggedright]{caption}

\begin{document}

\begin{center}
{\Large\textbf {Target shape dependence in a simple model of receptor-mediated endocytosis and phagocytosis}}
\\
\vspace{2mm}
David M. Richards$^{1,2,*}$,
Robert G. Endres$^{2}$
\\
\vspace{2mm}
{\scriptsize
$^1$ Wellcome Trust Centre for Biomedical Modelling and Analysis, Exeter University, Exeter EX2 5DW, UK\\
$^2$ Department of Life Sciences, Imperial College, London SW7 2AZ, UK\\
$^*$ E-mail: david.richards@exeter.ac.uk
}
\end{center}

\begin{abstract}
Phagocytosis and receptor-mediated endocytosis are vitally important particle uptake mechanisms in many cell types, ranging from single-cell organisms to immune cells. In both processes, engulfment by the cell depends critically on both particle shape and orientation. However, most previous theoretical work has focused only on spherical particles and hence disregards the wide-ranging particle shapes occurring in nature, such as those of bacteria. Here, by implementing a simple model in one and two dimensions, we compare and contrast receptor-mediated endocytosis and phagocytosis for a range of biologically relevant shapes, including spheres, ellipsoids, capped cylinders, and hourglasses. We find a whole range of different engulfment behaviors with some ellipsoids engulfing faster than spheres, and that phagocytosis is able to engulf a greater range of target shapes than other types of endocytosis. Further, the 2D model can explain why some nonspherical particles engulf fastest (not at all) when presented to the membrane tip-first (lying flat). Our work reveals how some bacteria may avoid being internalized simply because of their shape, and suggests shapes for optimal drug delivery.
\end{abstract}


\section*{Introduction}

Cells are capable of ingesting a huge range of particle shapes, from rod-shaped {\it E. coli} to doubly-lobed budding yeast, from helical {\it Borrelia} to filamentous {\it Legionella} \cite{HourglassII,Borrelia,Legionella}. Further, the shape is critical in determining whether engulfment is possible and, if so, how long it takes. In fact, it has been argued that the target shape plays an even more critical role than size \cite{ChampionI,ChampionII}. In addition, target orientation strongly affects internalisation, with ellipsoidal particles engulfed far easier when the highly-curved tip is presented first \cite{ChampionI,Howard}. Yet despite this, there is relatively little work studying shape and orientation dependence, with the vast majority of previous papers, both experimental and theoretical, involving only spherical targets.

Endocytosis encompasses a number of different mechanisms of cellular internalisation, including clathrin-mediated endocytosis, caveolar-type endocytosis, macro-pinocytosis and phagocytosis \cite{Doherty2009}. Whilst sharing some similarities, these processes often proceed in strikingly different manners. For example, phagocytosis is typically highly active and involves the membrane extending outwards, with finger-like protrusions surrounding the target in an actin-dependent process \cite{SwansonReceptorDynamics,DartMyosin1G}. Conversely, clathrin-mediated endocytosis is a more passive process, with targets appearing to sink into the cell \cite{Smythe1991}. Here we focus on types of endocytosis that involve target recognition via receptors. We assume that receptors bind irreversibly to ligands on the target so that engulfment proceeds monotonically \cite{GriffinI,GriffinII}. In particular, we distinguish phagocytosis from other less active forms of receptor-driven internalisation. We refer to all these latter processes as receptor-mediated endocytosis, including clathrin-mediated endocytosis.

Various mathematical models have attempted to understand the mechanism and dynamics of engulfment. All forms of endocytosis (in particular phagocytosis) are extremely complex, involving hundreds of different protein species in signalling cascades and cytoskeleton remodelling \cite{Samaj2004,Qualmann2002,SwansonSignallingI,UnderhillOzinsky}. As a result, overly-complicated models, that try to include every component, are unlikely to be useful. Instead more progress can be made by considering simpler, intuitive models that capture the essential mechanisms. For example, viral engulfment has been modelled by examining the free energy of membrane bending and adhesion \cite{Sun2005,Tzlil2004}. Similarly van Effenterre et al. used a thermodynamic approach that involved considering an ensemble of target particles \cite{Effenterre2003}. Recently, the role of actin during endocytosis in yeast was addressed by using a variational approach and arguing that the final pinch-off stage is due to a pearling-like instability \cite{Zhang2015}.

There are also models that focus on phagocytosis. For example, van Zon et al. included simple dynamics for both actin and receptors, which they used to understand why engulfment normally either stalls before halfway or reaches completion \cite{Howard}. Herant et al. focussed on the forces required to explain the cup shape and showed, using a continuum mechanics approach, the need for both repulsion at the cup edge and flattening within the cup \cite{Heinrich1B,HeinrichModel2}. Various other approaches focus only on energetic requirements, such as those due to membrane bending and receptor-ligand binding, equating the phagocytic cup with some minimum energy state. For example, Dasgupta et al. used this approach to argue that ellipsoids are harder to engulf than spheres \cite{DasguptaI,DasguptaII}, whereas Tollis et al. found that an actin-driven ratchet mechanism can lead to robust engulfment \cite{TollisZipper}.

One particularly elegant approach by Gao et al. models endocytosis by considering only the motion of receptors within the membrane \cite{Freund}. They argued that the essence of receptor-mediated endocytosis is related to the dynamics of the receptors themselves, which can be mapped to the supercooled Stefan problem, a simple physical model of how the boundary between ice and water moves during freezing. Although they were able to understand how particle radius affects the rate of engulfment (predicting an optimum radius corresponding to the quickest possible engulfment time), they only considered spherical particles. Cylindrical ellipsoids were considered in \cite{DecuzziFerrari}, but not in a consistent manner. In \cite{OurPaper} we extended the Gao et al. model to phagocytosis, arguing that a similar mathematical model can explain receptor motion in both cases. This involved including receptor drift and signalling, and showed that the two distinct stages of engulfment (an initial slow stage followed by a much quicker second stage) can be explained by different receptor dynamics during different stages. However, again, we almost exclusively focused on only spherical particles.

Here, to address the role of particle shape during engulfment, we extend these models in three ways. First, by allowing nonconstant curvature, we are able to study nonspherical particle shapes, such as ellipsoids, capped cylinders and hourglasses (Fig. 1). All the shapes we consider are relevant for real biological systems, such as rod-shaped bacteria and dividing budding yeast. Second, we include signalling and a role for actin in order to examine the different target shape dependence between phagocytosis and other types of receptor-mediated endocytosis. Third, we extend the formalism from a 1D to a 2D model, which allows, for the first time to our knowledge, lower-symmetry shapes, such as lying ellipsoids, to be studied. With this model we are able to address how orientation affects engulfment, finding an explanation for why prolate spheroids engulf quickest when presented to the cell tip-first. Finally, we compare our model with known experimental results.


\section*{Results}

We first study endocytosis in general and consider particles both with and without circular symmetry. Then we extend the model to apply to phagocytosis by including signalling and actin. By particle symmetry we mean the symmetry of the particle looking from directly above the membrane. For example, a prolate spheroid will have circular symmetry only when presented tip first (as in the shape on the left in Fig. 1B). As we now show, our model for circularly-symmetric particles becomes effectively 1D, allowing easier analytic and numerical treatment. Conversely, non-symmetric particles require a full 2D model, introducing extra physical and numerical issues. We first consider circularly-symmetric particles---spheres, tip-first spheroids, capped cylinders, hourglasses---and next examine non-symmetric particles, for example ellipsoids that lie flat (as in the first shape in Fig. 1B lying on its side).

\begin{figure}[t]
  \centerline{\includegraphics[width=0.6\textwidth]{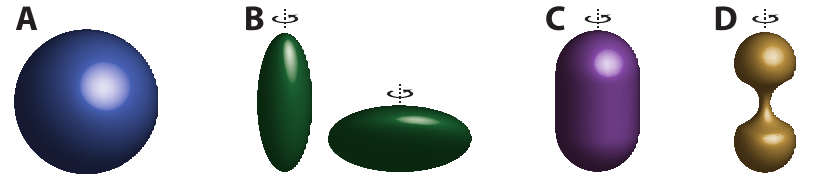}}
  \caption{{\bf Biologically relevant particle shapes.} (A) Spheres. (B) Spheroids, both prolate (left) and oblate (right). (C) Capped cylinders, consisting of a cylindrical region capped by hemispheres, which describe certain bacteria. (D) Hourglasses made from spherical caps and a narrow neck region, which model cells during the division process.}\label{Fig1}
\end{figure}


\subsection*{1D model for endocytosis of circularly-symmetric particles}

We first extend the model in Gao et al. \cite{Freund} to include nonspherical particles. This is a model of receptor motion with the inclusion of membrane bending and ligand-receptor binding, which focuses on the cup size as a function of time. Consider a particle whose cross section parallel to the membrane is circularly symmetric (such as all the shapes in Fig. 1). Then receptor motion within the membrane and the fraction of engulfment can be parameterised by just one spatial parameter, $r$, the distance from the centre of the cup. Thus, at time $t$, the receptors are described by a density $\rho(r,t)$ and the engulfment by $a(t)$, which represents the engulfed arc length measured from the centre of the cup (Fig. 2A).

Initially, $\rho$ is assumed uniform and given by $\rho_0$, which is also the value that is imposed at $r=\infty$ at all times. As the particle is engulfed, receptors bind to ligands on the bead and the density increases to $\rho_L$ in the engulfed region (Fig. 2B). We assume for simplicity that receptors are never destroyed or created, so that the evolution of the receptor density, $\rho$, in the non-engulfed region ($r\ge a$) is governed simply by diffusion with diffusion constant $D$. At the cup boundary receptors flow into the bound region, increasing the engulfment $a$. By considering the conservation of receptors, the rate of increase of $a$ can be related to the flux of receptors, so that our 1D model is given in the non-engulfed region $r\ge a$ by
\begin{equation} \label{eq:1D_eqs}
  \frac{\partial\rho}{\partial t} = \frac{D}{r} \frac{\partial}{\partial r}\left( r\frac{\partial\rho}{\partial r} \right), \qquad \frac{da}{dt} = \frac{D\rho'_+}{\rho_L-\rho_+},
\end{equation}
where $\rho_+$ is the receptor density at the cup edge. Initial conditions are $\rho(r,0)=\rho_0$ and $a(0)=0$. This is identical to the supercooled 1D Stefan problem, studied in physics as a model of the freezing of water.

To find a unique solution we impose one extra condition, the value of $\rho$ at the edge of the cup, $\rho_+$. Since the boundary continually moves, the position where this extra condition is applied also continually moves, such that $\rho_+(t)=\rho(a(t),t)$. As in \cite{Freund}, we fix $\rho_+(t)$ by requiring that there is no free-energy jump across the cup, {\it i.e.} that all energy from receptor-ligand binding is used for engulfment. We consider three contributions to the free energy (receptor-ligand binding, membrane curvature and receptor entropy), which leads to the condition
\begin{equation} \label{eq:power_bal}
  \frac{\rho_+}{\rho_L} - \ln\left(\frac{\rho_+}{\rho_L}\right) = \mathcal{E} - \frac{2\mathcal{B}H^2}{\rho_L} + 1,
\end{equation}
where $\mathcal{E}$ is the binding energy per receptor-ligand bond, $\mathcal{B}$ is the bending modulus, and $H$ is the mean curvature of the particle at the cup edge. We ignore any contribution from the Gaussian curvature since its effect is normally similar to that from the mean curvature for biologically relevant target shapes.

The original model in \cite{Freund} only considered spherical particles, so that the curvature $H$ is constant. In order to consider other particle shapes, we allow $H$ to depend on the engulfment $a$. This in turn means that $\rho_+$ is now not constant in time and that analytic solutions can no longer be found. This same approach was attempted in \cite{DecuzziFerrari} for cylindrical particles, but with a misunderstanding that we believe invalidates their results. To make progress, we numerically solve the system (see Materials and Methods) and consider, in turn, spheroids, capped cylinders and hourglasses.

\begin{figure}[t]
  \centerline{\includegraphics[width=0.6\textwidth]{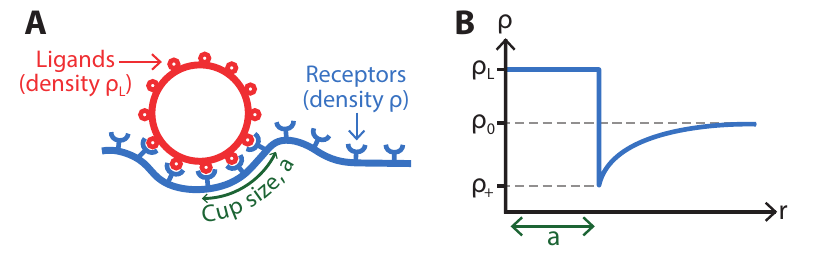}}
  \caption{{\bf Sketch of the model.} (A) Sketch of the receptors (blue) moving within the membrane and a partially engulfed particle (red). The model variables are the cup size ($a$) and the receptor density ($\rho$). (B) The typical profile of $\rho$, with $\rho_L$ the ligand density, $\rho_0$ the receptor density at infinity, and $\rho_+$ the receptor density at the cup edge.}\label{Fig2}
\end{figure}


\subsection*{Spheroids}

A spheroid is an ellipsoid where two of the principal axes have the same length. Later, in our 2D model, we will be able to consider spheroids presented to the cell both end-on and lying down. However, in the 1D model, which requires circularly-symmetric particles, only standing spheroids can be examined. A spheroid is characterised by two radii: $R_1$, the radius of the semi-principal axis parallel to the membrane, and $R_2$, the radius perpendicular to the membrane (Fig. 3A). We use a standard parameterisation of the surface of a spheroid given by $(x,y,z) = (R_1 \sin v \cos u, R_1 \sin v \sin u, -R_2 \cos v)$, where $u\in[0,2\pi)$ and $v\in[0,\pi]$. Here the $z$-axis represents the perpendicular distance from the membrane, so that $v=0$ is the ``lowest'' point, the point of the spheroid that first touches the membrane and is the first to be engulfed.

We must relate the mean curvature $H$ to the arc length $s$ (measured from $v=0$). The curvature can be written in terms of $v$ as
\begin{equation} \label{eq:spheroid_curvature}
  2H = \frac{R_2\left( R_1^2(1+\cos^2\!v) + R_2^2\sin^2\!v \right)}{\displaystyle R_1\left( R_1^2\cos^2\!v + R_2^2\sin^2\!v \right)^{3/2}},
\end{equation}
whereas the arc length is given by an incomplete elliptic integral of the second kind,
\begin{equation} \label{eq:spheroid_arc_length}
  s = \int_0^v\sqrt{\displaystyle R_1^2\cos^2\!w + R_2^2\sin^2\!w }\,dw.
\end{equation}
Since $v$ cannot be directly eliminated between Eqs. \eqref{eq:spheroid_curvature} and \eqref{eq:spheroid_arc_length}, we proceed as follows: for a given engulfment (arc length) $a$ we numerically solve Eq. \eqref{eq:spheroid_arc_length} to find $v$ and then use Eq. \eqref{eq:spheroid_curvature} to find the curvature.

We find that, for $R_1\neq R_2$, engulfment no longer proceeds with the square root of time (Fig. 3B). The speed of engulfment now depends on the local curvature at the cup edge. Higher curvatures decrease the right hand side of Eq. \eqref{eq:power_bal} and increase $\rho_+$. In turn a higher $\rho_+$ decreases $\rho'$ at the cup edge and tends to reduce the engulfment rate. Thus highly curved regions lead to slow engulfment, whereas flatter regions engulf faster. Regions which are too highly curved lead to stalling, where engulfment cannot proceed further (Fig. 3B). In our mathematical model this occurs when $\rho_+\ge\rho_0$ so that the right-hand side of the second part of Eq. \eqref{eq:1D_eqs} becomes negative, which we interpret as $\dot{a}=0$.

To compare different spheroids, we vary $R_1$ whilst always choosing $R_2$ to ensure the same total surface area. We find similar results by instead fixing the volume or $R_2$. As shown in Figs. 3B and 3C, we find five classes of behaviour, depending on the value of $R_1$. First, for sufficiently small $R_1$, the initial curvature is so high that engulfment can never begin and the particle simply sits on the cell membrane. Second, for larger $R_1$ we reach a slow-fast-slow mode of engulfment: the initial high curvature at the particle base leads to slow engulfment, with quicker engulfment in the smaller-curvature middle region, before again a highly-curved, slow stage near the top of the particle. Third, as $R_1$ is further increased, the spherical case is reached ($R_1=R_2$), characterised by $\sqrt{t}$ engulfment. Fourth, at still higher $R_1$, a fast-slow-fast stage is encountered, with engulfment quickest at the beginning and end where the curvature is lowest. Finally, sufficiently high $R_1$ again leads to incomplete engulfment, although now (unlike for small $R_1$) engulfment stalls around halfway.

To better understand the effect of changing $R_1$, we study the half and full engulfment times (Fig. 3D). Spheroids with $R_1<R_2$ (like those on the left in Fig. 1B) always take longer than the equivalent sphere to reach both half and full engulfment. Conversely, spheroids with $R_1$ slightly bigger than $R_2$ engulf faster than spheres. This is interesting and suggests that rod-shaped bacteria can sometimes be engulfed faster than coccus-shaped bacteria of the same surface area. As $R_1$ is further increased a minimum is reached. Beyond this the high curvature becomes a dominant factor and causes both the half and full engulfment times to rise sharply for larger values of $R_1$.

To study the orientation dependence of engulfment, we compare ellipsoids with $\{R_1,R_2\}=\{R,1\mu\rm{m}\}$ to those with $\{R_1,R_2\}=\{1\mu\rm{m},\tilde{R}\}$, where $\tilde{R}$ is chosen so that the surface area is the same for both cases. Of course, these two shapes do not correspond to the same spheroid; only later, in our extended 2D model, we will be able to directly compare the same ellipsoid in different orientations. In particular we compare the difference in total engulfment time for these two shapes (Fig. 3E). For $R>1\mu\rm{m}$ we find that the oblate spheroid (with $\{R_1,R_2\}=\{R,1\mu\rm{m}\}$) engulfs quickest. This is simply because the total engulfment length is smaller. For the same reason, for $R$ slightly smaller than $1\mu\rm{m}$, engulfment is also quickest for oblate spheroids (which are now those with $\{R_1,R_2\}=\{1\mu\rm{m},\tilde{R}\}$). However, for sufficiently small $R$, the required value of $\tilde{R}$ is such that the curvature at the mid-point becomes limiting, slowing engulfment for the $\{R_1,R_2\}=\{1\mu\rm{m},\tilde{R}\}$ spheroid, so that prolate spheroids are engulfed faster.

\begin{figure}[!ht]
  \centerline{\includegraphics[width=0.6\textwidth]{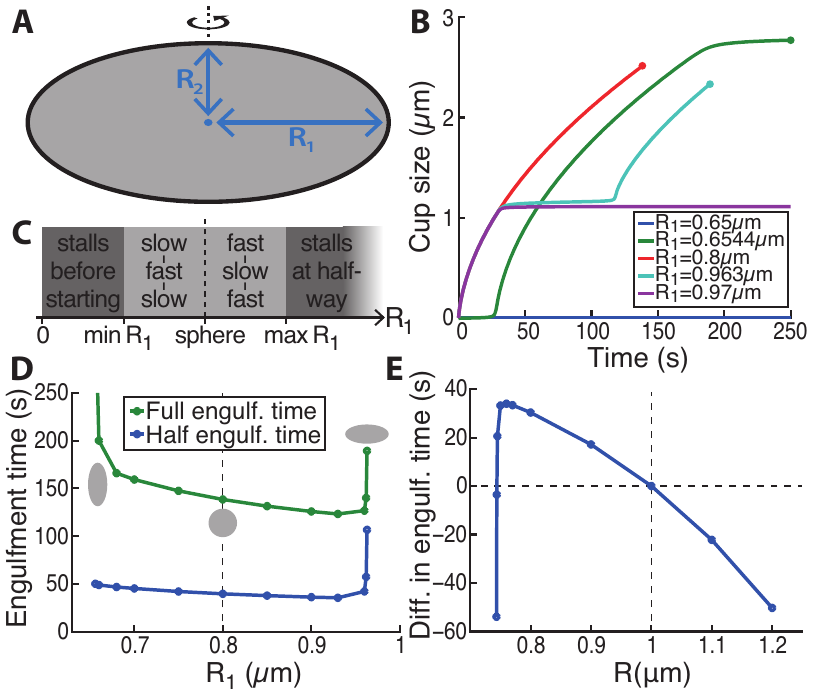}}
  \caption{{\bf Engulfment time for spheroids in 1D model.} (A) Cross section of a spheroid, defined by $R_1$ (the axis length parallel to the membrane) and $R_2$ (the perpendicular axis length). The full shape is obtained by rotating around the z-axis, thus giving an ellipsoid with axis lengths $\{R_1,R_1,R_2\}$. (B) Progression of engulfment for spheroids with various $R_1$. In each case $R_2$ is chosen so that the surface area is the same as for a sphere of radius $0.8\mu\rm{m}$. The spherical case (red curve) shows classic $\sqrt{t}$ engulfment. Nonspherical spheroids show slower rates of engulfment in regions with high curvatures. This occurs at the beginning and end of engulfment for prolate spheroids (green curve) and towards the middle for oblate spheroids (cyan curve). Sufficiently eccentric spheroids (blue and purple curves) stall at the point where the curvature gets so high that $\rho_+>\rho_0$. Solid dots represent complete engulfment. (C) The five types of behaviour for increasing $R_1$ (for fixed surface area): (i) stalling before engulfment starts, (ii) slow-fast-slow engulfment, (iii) classic $\sqrt{t}$ spherical behaviour, (iv) fast-slow-fast engulfment, and (v) stalling at halfway. (D) The half- and full-engulfment times as a function of $R_1$. Again $R_2$ is chosen to keep the surface area constant. The dashed line represents the spherical case. Interestingly, the quickest engulfment occurs, not for spheres, but for oblate spheroids (where $R_1>R_2$). Grey shapes: sketches of the target particle. (E) The difference in total engulfment time between ellipsoids with radii $\{R,R,1\mu\rm{m}\}$ and those with $\{1\mu\rm{m},1\mu\rm{m},\tilde{R}\}$, where $\tilde{R}$ is chosen to give the same surface area. The vertical dashed line represents the case of a sphere. Parameters: $\rho_0=50\mu\rm{m}^{-2}$, $\rho_L=500\mu\rm{m}^{-2}$, $D=0.4\mu\rm{m}^2\rm{s}^{-1}$, $\mathcal{E}=15$, $\mathcal{B}=10$.}\label{Fig3}
\end{figure}


\subsection*{Capped cylinders and hourglasses}
We now briefly consider two other circularly-symmetric shapes. Capped cylinders, consisting of a cylinder with a hemisphere attached at either end, are good models for various rod-shaped bacteria, such as {\it E. coli} and {\it Bacillus subtilis}. These display a slow-fast-stall-slow engulfment behaviour, related to the lower curvature found in the neck region. The stalling is caused by a jump in $\rho_+$ at the top of the neck. By fixing the surface area and varying the neck height, we find that the quickest engulfment corresponds to a sphere. Even though rods are not capped cylinders, this could explain the observation that, for endocytosis, spheres engulf in less time than rod-shaped particles \cite{Chithrani2006}.

Hourglasses are more complicated shapes, but are useful for modelling cases of dividing cells, such as budding yeast during division \cite{HourglassII,Hourglass}. The engulfment behaviour and the total engulfment time depend on exactly how narrow the neck is compared to the spherical caps, with the neck region engulfed sometimes faster and sometimes slower than the caps. Interestingly, the quickest engulfed hourglass does not correspond to the shape with the smallest neck curvature.


\subsection*{Extension to phagocytosis}

Although actin plays a role in many types of endocytosis, it seems to be far more important for successful engulfment in phagocytosis \cite{Engqvist2003,Robertson2009,TollisZipper}. To this end, we extend our model to apply to phagocytosis by adding a simple role for actin. Motivated by our previous work \cite{OurPaper}, we achieve this by including a signalling molecule $S$ which recruits actin to push at the cup edge. We assume that $S$ is produced only within the cup (with rate $\beta\rho_L$), is degraded everywhere (with lifetime $\tau$), and diffuses (with constant $D_S$). Then $S$ is described by \cite{OurPaper}
\begin{align}
  \label{eq:sig_eq}
  \frac{\partial S}{\partial t} = \frac{D_S}{r} \frac{\partial}{\partial r}\left( r\frac{\partial S}{\partial r} \right) + \beta\rho_L\Theta(a-r) - \tau^{-1}S,
\end{align}
where $\Theta(x)$ is the Heaviside function.

Even though the actin network provides a pushing force at the cup edge, this will not necessarily increase the engulfed target length, $a$, unless sufficient receptors are also present: simply pushing the membrane further around the target without a suitable zipper mechanism risks subsequent unwrapping \cite{TollisZipper}. Thus the real effect of actin (at least as far as our model is concerned) is in facilitating receptor motion to the cup edge. One way this could be achieved is by actin helping to bend the membrane around the target, thus effectively reducing the free energy needed for membrane bending. For example, an actin network structure that matches the target curvature would naturally encourage the membrane to bend in the same manner via scaffolding \cite{McMahon2015}. We implement this mechanism by allowing the bending modulus, $\mathcal{B}$, to depend on the value of $S$ at the cup edge, $S_+$, via $\mathcal{B} = \mathcal{B}_0/(1+\mathcal{B}_1S_+)$, where $\mathcal{B}_i$ are constants. We identify $\mathcal{B}_0$ with the (constant) value of $\mathcal{B}$ in the non-phagocytic model and call $\mathcal{B}_1$ the bending-actin constant. Then $\mathcal{B}_1=0$ corresponds to non-phagocytic endocytosis and $\mathcal{B}_1>0$ to phagocytosis. Note that this mechanism does not require that the membrane becomes softer, only that bending becomes easier. This could be related, for example, to changes in lipid composition within the phagocytic cup as a result of localised small GTPase recruitment \cite{Hoppe2004} or to changes in the spontaneous curvature.

We first examine the difference between receptor-mediated endocytosis and phagocytosis for oblate spheroids ($R_1>R_2$). Sufficiently eccentric spheroids cannot be engulfed by receptor-mediated endocytosis since the curvature around half way is too high (Fig. 3D). However, with the introduction of signalling and a role for actin, such ellipsoids can often be ingested by phagocytosis (Fig. 4A). In these cases, the initial rate of engulfment, when $S$ is small, is similar in both models. However, when engulfment slows down around half way, the recruitment of actin to the cup edge provides an extra push. If the bending-actin constant is large enough this then leads to complete engulfment (Fig. 4A). The situation is markedly different for prolate spheroids ($R_1<R_2$) for which the high curvature stalling occurs at the beginning of engulfment when there are relatively few receptor-ligand bonds. The lack of bound receptors means that there is little signalling and so little recruitment of actin. In such cases, often both models fail to engulf.

For oblate spheroids we can examine the effect of the bending-actin constant on engulfment time in cases where non-phagocytic endocytosis stalls around half way (Fig. 4A). Initially, as $\mathcal{B}_1$ is increased, there is still no engulfment. However, at a critical onset (about $\mathcal{B}_1=0.315\mu\rm{m}^2$ with the parameters in Fig. 4A) complete engulfment occurs. As $\mathcal{B}_1$ is increased still further, the engulfment time rapidly drops. For values near the critical $\mathcal{B}_1$ value, engulfment takes a long time, with the majority of time spent around half engulfment waiting for enough actin to be recruited to the cup edge. This suggests that certain shapes that appear to stall around half way will actually fully ingest given a sufficient amount of time.

Finally, we examine how the range of ingestible spheroids changes as the bending-actin constant increases (Fig. 4B). For any fixed value of $\mathcal{B}_1$ there is always a limit on which ellipsoids can be engulfed: particles that are too oblate or too prolate will never be ingested. However, these limits correspond to progressively more and more eccentric spheroids as $\mathcal{B}_1$ increases. From experimental measurements of which ellipsoids can and cannot be engulfed, it should be possible to infer the value of the bending-actin constant. Thus, phagocytosis ($\mathcal{B}_1>0$) allows not only larger particles, but also more eccentric particles to be ingested compared to other types of receptor-mediated endocytosis ($\mathcal{B}_1=0$). However, our model suggests that there are always sufficiently eccentrically shaped targets that can evade engulfment.

\begin{figure}[t]
  \centerline{\includegraphics[width=0.6\textwidth]{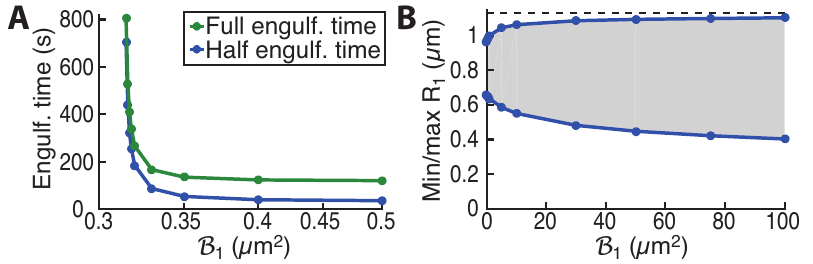}}
  \caption{{\bf Results for spheroids in phagocytosis model.} (A) Engulfment times (for both half and full engulfment) as a function of the bending-actin constant ($\mathcal{B}_1$) for a spheroid with $R_1=0.98\mu\rm{m}$ and $R_2=0.45\mu\rm{m}$. $\mathcal{B}_1=0$ corresponds to non-phagocytic endocytosis and $\mathcal{B}_1>0$ to phagocytosis. (B) The range of spheroids (grey region) that can be engulfed as a function of the bending-actin constant. $R_2$ is always chosen so that the surface area is the same as that of a $0.8\mu\rm{m}$-radius sphere. The dashed line represents the largest possible value of $R_1$: above this the surface area constraint can no longer be imposed. Parameters: $\rho_0=50\mu\rm{m}^{-2}$, $\rho_L=500\mu\rm{m}^{-2}$, $D=0.4\mu\rm{m}^2\rm{s}^{-1}$, $\mathcal{E}=15$, $\mathcal{B}_0=10$, $D_S=1\mu\rm{m}^2\rm{s}^{-1}$, $\beta=0.4\rm{s}^{-1}$, $\tau=0.5\rm{s}$.}\label{Fig5}
\end{figure}


\subsection*{2D model for noncircularly-symmetric particles}

A limitation of the 1D model is that only circularly-symmetric particles can be studied. In particular, this excludes general ellipsoids (with all the axes different lengths) and even spheroids that are not presented tip-first. To study such shapes we extend the above model to two dimensions, which involves studying a 2D Stefan problem. We parameterise the surface by polar coordinates $r$ and $\theta$, with $r\ge0$ and $0\le\theta<2\pi$. The receptor density is now described by $\rho(r,\theta,t)$, whereas the engulfed arc length, $a(\theta,t)$, becomes a function of the angular direction. With the same boundary conditions as in the 1D model ($\rho=\rho_0$ both at $t=0$ and at $r=\infty$), we now have to solve the 2D diffusion equation,
\begin{equation} \label{eq:2D_eqs_a}
  \frac{\partial\rho}{\partial t} = \frac{D}{r} \frac{\partial}{\partial r}\left( r\frac{\partial\rho}{\partial r} \right) + \frac{D}{r^2}\frac{\partial^2\rho}{\partial\theta^2},
\end{equation}
for $r>a$. As before, $\rho=\rho_L$ within the cup. The evolution of $a$ can be derived from the conservation of receptors, which leads to the second part of Eq. \eqref{eq:1D_eqs} being replaced by
\begin{equation} \label{eq:2D_eqs_b}
  \frac{\partial a}{\partial t} = \frac{D(\boldsymbol{\nabla}\rho_+\cdot\boldsymbol{\hat{n}})}{\rho_L-\rho_+},
\end{equation}
where $\boldsymbol{\hat{n}}$ is a unit outward-pointing normal to the cup surface. The final boundary condition, fixing $\rho_+$, again follows from requiring no free-energy jump across the cup edge. Although the derivation is more involved, the final condition is the same as in the 1D model, Eq. \eqref{eq:power_bal}. To find solutions we developed a numerical approach that involves a 2D lattice that must be continually updated as the cup enlarges (see Materials and Methods). To test our numerics we checked that spherical particles (with constant curvature $H=R^{-1}$) agree with the analytic results from the 1D model.

We focus on studying ellipsoids, which are parameterised by their three semi-principal axes, $R_1$, $R_2$ and $R_3$ (Fig. 5A). Since the cup does not proceed in unison around the particle (different angles engulf at different rates), there is now no unique definition of half engulfment. We consider two definitions: half-area engulfment when the engulfed area is half the ellipsoid surface area, and half-circumference engulfment when all points of the cup are over half way to the top of the particle. Half-area engulfment always occurs before (or at the same time as) half-circumference engulfment.

As an example of engulfment, consider an ellipsoid with $R_1=R_3<R_2$, which is a lying-down prolate spheroid as in the first figure in Fig. 1B if it was lying on its side (see Video S1). The lack of circular symmetry (when viewed from above) means that engulfment now proceeds at different rates at different angles. In particular, consider $\theta=0$ compared to $\theta=\frac{\pi}{2}$, corresponding to the two principal axes parallel to the membrane. Cup progression at $\theta=0$ proceeds along a circle of radius $R_1$ with constant curvature, $H_{\theta=0}$. Conversely, the cup edge at $\theta=\frac{\pi}{2}$ follows an ellipse, with the curvature starting at $H_{\theta=0}$, rising to a maximum at half-way and then decreasing back to $H_{\theta=0}$. In addition, the total arc length along $\theta=\frac{\pi}{2}$ is longer than that along $\theta=0$, so that the $\theta=\frac{\pi}{2}$ direction must engulf more membrane to reach full engulfment.

As seen in Fig. 5B and Video S1, the $\theta=\frac{\pi}{2}$ direction takes considerably longer to engulf. This is partly due to both the greater curvature along $\theta=\frac{\pi}{2}$ and the greater arc length. However, there is another, more important reason that slows engulfment at $\theta=\frac{\pi}{2}$ compared to $\theta=0$, which is only revealed in the full 2D simulation. Once $\theta=\frac{\pi}{2}$ starts to lag behind $\theta=0$ (due to higher curvature), angular diffusion (diffusion in the $\theta$ direction) causes even more receptors to move away from $\theta=\frac{\pi}{2}$ and towards $\theta=0$. This increases the flux at $\theta=0$, so that the $\theta=\frac{\pi}{2}$ direction slips even further behind $\theta=0$, amplifying the effect. This amplification explains why engulfment takes so much longer at $\theta=\frac{\pi}{2}$. This effect may also explain the asymmetric spreading over ellipsoidal particles in \cite{ChampionI}.

For the first time, to our knowledge, we are now able to compare spheroids presented to the cell with different orientations. To do this we compare spheroids with radii $\{R_1,R_2,R_3\}=\{\tilde{R},\tilde{R},R\}$ to those with $\{R_1,R_2,R_3\}=\{\tilde{R},R,\tilde{R}\}$. We refer to the first case as a symmetrically-presented spheroid (since from above this spheroid has circular symmetry) and the second case as an asymmetrically-presented spheroid. For a given $R$ we always choose $\tilde{R}$ to keep constant surface area (the surface area of a sphere with radius $0.4\mu\rm{m}$). This allows us to directly compare spheroids with different eccentricities. Fig. 5C shows the results for the total engulfment time. Both symmetrically- and asymmetrically-presented spheroids demonstrate the expected characteristic behaviour, with quickest engulfment corresponding to some intermediate spheroid: more squashed or more pointed ellipsoids take longer to engulf, with sufficiently eccentric shapes never reaching complete engulfment. Further, the symmetrically-presented spheroid always engulfs first, both for prolate and oblate spheroids. This is due to the amplification of angular diffusion described above, where the $\theta=\frac{\pi}{2}$ angle for asymmetrically-presented spheroids takes much longer to engulf than $\theta=0$.

For symmetrically-presented spheroids we find, as we should, similar results to those for spheroids in the 1D model. For fixed surface area, the sphere is not the optimal shape. Rather slightly oblate spheroids ($R<\tilde{R}$) engulf faster. This is true not only for the total engulfment time, but also for the half-area and half-circumference engulfment times (which are always equal for such a shape). Conversely, for asymmetrically-presented spheroids, the sphere always corresponds to quickest total engulfment. This again is caused by the amplification effect, which penalises non-symmetric shapes. The half-circumference engulfment time (which requires all angles to have reached half-engulfment) is also a minimum for spheres (Fig. S5B). However, the situation is different for the half-area engulfment time, where prolate spheroids ($R>\tilde{R}$) reach half-area engulfment sooner than spheres (Fig. S5A).

Thus our 2D model leads to three important conclusions for spheroids. First, prolate spheroids engulf quickest when presented to the membrane tip-first, which agrees with previous phagocytic measurements \cite{ChampionI} and is a prediction for other types of endocytosis. Sufficiently spherical targets will completely engulf in either orientation, whereas targets that are too eccentric are not engulfed in any orientation. This could well explain the observation in \cite{ChampionI}, where the membrane sometimes only spreads along the target with no internalisation. However, there also exists an intermediate regime where prolate spheroids are engulfed only when one of the highly-curved tips is engulfed first. Although in phagocytosis this orientation dependence can be explained by adding active processes and a second engulfment stage to the 1D model \cite{OurPaper}, our 2D model shows that this (for all types of endocytosis) can also be understood simply by receptor diffusion. Second, we predict that oblate spheroids engulf quickest when presented flat to the membrane. This finding is perhaps surprising since it is then the low curvature region that is engulfed first (and last) rather than the highly-curved tips as for prolate spheroids. Hence oblate and prolate spheroids have completely different engulfment behaviours even for superficially similar shapes (such as the two standing spheroids in Fig. 5C). Further experiments with spheroidal targets should help clarify this prediction. Third, the 2D model predicts that the absolute quickest engulfment for a spheroid (with fixed surface area) occurs for a slightly oblate spheroid that lies flat with respect to the membrane. The last finding is confirmed by Sharma et al. for macrophages \cite{Sharma2010}.

\begin{figure}[t]
  \centerline{\includegraphics[width=0.6\textwidth]{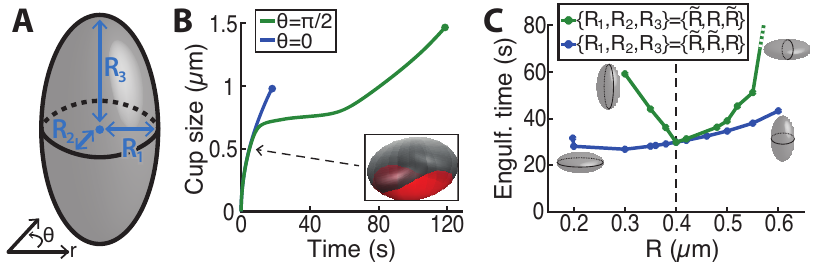}}
  \caption{{\bf Engulfment time for ellipsoids in 2D model.} (A) Sketch of an ellipsoid, which is defined by its three semi-principal axis lengths, $R_1$, $R_2$ and $R_3$. (B) Comparison of engulfment at different angles for an ellipsoid with $R_1=R_3=0.3\mu\rm{m}$ and $R_2=0.6\mu\rm{m}$. Different directions around the ellipsoid engulf at different rates, with complete engulfment occurring just after $118\rm{s}$. Here we show the two most extreme angles: $\theta=0$ and $\theta=\frac{\pi}{2}$. Insert: cup shape when the $\theta=0$ direction is about half engulfed with the target in grey and the cell membrane in red. For the full movie, see Video S1. (C) Comparison of total engulfment time for spheroids presented to the cell in two perpendicular orientations. $\tilde{R}$ is always chosen so that the surface area is the same as that of a sphere of radius $0.4\mu\rm{m}$. The dashed line represents the spherical case. Grey shapes: sketches of the target particle. Parameters: $\rho_0=50\mu\rm{m}^{-2}$, $\rho_L=500\mu\rm{m}^{-2}$, $D=0.4\mu\rm{m}^2\rm{s}^{-1}$, $\mathcal{E}=6$, $\mathcal{B}=0.4$.}\label{Fig6}
\end{figure}


\section*{Discussion}

In analogy with the freezing of water, as described by the heat equation with a moving boundary, we used a minimal model of receptor diffusion and capture to study shape and orientation dependence during endocytosis. The process of engulfment is then equivalent to a first-order phase transition characterised by a transition from mobile, unbound receptors to immobile, ligand-bound receptors. By including signalling and a role of actin, we were then able to adapt our model to phagocytosis and examine the different role shape plays for receptor-mediated endocytosis and phagocytosis.

We both extended a previous 1D model to include nonspherical targets and developed a more realistic 2D version, allowing lower-symmetry particles to be studied. The 1D model shows that a whole host of engulfment behaviours emerge from different particle types. This supports the idea that the local curvature at the cup edge is a crucial factor in determining whether (and how quickly) targets are engulfed \cite{ChampionI}. Highly-curved regions may prevent complete engulfment and so are likely to be factors that, at least partially, drive the evolution of pathogen shapes. We found that for some shapes (such as capped cylinders) the more spherical the target the quicker the engulfment, whereas for other shapes (such as spheroids) nonspherical particles can be engulfed faster. Thus long, narrow filamentous shapes could in some situations be more beneficial for pathogens that try to avoid ingestion. Further, our hourglass-shaped particles also explain the observation that phagocytosis of budded yeast can stall at half engulfment \cite{HourglassII,Hourglass}.

The 2D model allowed, for the first time to our knowledge, the orientation dependence of engulfment to be probed. We found that, for nonspherical targets, engulfment can sometimes proceed at completely different rates in different directions, with higher-curved directions engulfing much slower. This should be directly observable with, for example, scanning electron microscopy. Further, we discovered that this effect can explain the experimental observation that prolate spheroids phagocytose quickest when presented to the cell tip-first \cite{ChampionI}. In addition, the 2D model predicts both that oblate spheroids engulf fastest when presented flat to the cell and that slightly oblate spheroids can engulf even faster than the equivalent sphere. These perhaps counterintuitive results could readily be tested in the near future.

Although phagocytosis is distinct to other forms of receptor-mediated endocytosis, the role of receptors and their motion towards the cup is equally important \cite{OurPaper}. By adding signalling and a simple role for actin, our models predict that all types of endocytosis, including phagocytosis, have similar shape dependence. In addition we found that phagocytosis not only fulfils its traditional role of allowing larger particles to be ingested, but also allows more highly-curved targets to be engulfed.

It is important to point out the limitations of our models. We have focussed only on passive diffusion of receptors, neglecting more active receptor motion \cite{SwansonSignallingI,UnderhillOzinsky}. Further, in our phagocytosis model, we have only included one signalling molecule and have added actin in a highly-simplified manner without any attempt to model the underlying structure of the cytoskeleton. Although these simplifications will not capture many aspects of endocytosis, it is likely that the interplay between target (and hence membrane) curvature and the motion of receptors is correctly represented even in our minimal models. Our phagocytosis model also assumes that signalling reduces the membrane bending modulus, effectively making bending around highly curved objects easier. An alternative model representation could have focussed on forces instead of receptor dynamics so that, for example, signalling and actin polymerisation would lead to a membrane pushing force. Taking spatial derivatives of the cup energy would translate our energy-based model to a force-based model \cite{Howard}. Finally, we have assumed that targets do not rotate during engulfment. It would be interesting to also include this in our models, which would then allow further questions, such as the stability of different orientations, to be addressed.

Shape dependence is a fascinating aspect of endocytosis, which is probably even more important for successful engulfment than target size \cite{ChampionI}. Understanding the role of shape may also have direct applications to, for example, how some pathogens invade cells \cite{MycobacteriaEjection}, how other pathogens avoid uptake, the optimal design for drug delivery carriers \cite{DrugShapeI,DrugShapeII}, and even cell behaviour \cite{Huang2010}. We have shown that there is great variety in the uptake of different target shapes, which is also heavily orientation dependent. Future work, using differently coated particles, will need to address how geometric, mechanical and chemical signals are integrated to explain the wide range of observed engulfment outcomes.


\section*{Materials and Methods}

\subsection*{1D numerical simulations}
For nonspherical particles, even the 1D model must be solved numerically. We used a grid with spacing $\Delta r=0.1\mu\rm{m}$ and total length $L=50\mu\rm{m}$. Eq. \eqref{eq:1D_eqs} was solved using the Euler method with time step $\Delta t = 1\times10^{-3}\rm{s}$. At each time step we also imposed the boundary condition $\rho(a(t),t)=\rho_+$, which involves first calculating the mean curvature at the cup edge and solving Eq. \eqref{eq:power_bal}. We checked convergence by showing that smaller values of $\Delta r$ and $\Delta t$ did not noticeably change the results.

\subsection*{2D numerical simulations}
Our 2D numerical solutions were calculated on a 2D grid with $\Delta r=2.5\rm{nm}$ and $\Delta\theta=\pi/8$. The maximum radius was set at $20\mu\rm{m}$. The area of each lattice point was no longer constant, but depended on $r$, a fact that had to be taken into account. We solved Eqs. \eqref{eq:2D_eqs_a} and \eqref{eq:2D_eqs_b} with the usual Euler method using $\Delta t = 2.5\times10^{-6}\rm{s}$. At a given time $t$ and angle $\theta$, the normal to the cup surface was estimated from $a(\theta,t)$ and $a(\theta\pm\Delta\theta,t)$. Similarly, $\boldsymbol{\nabla}\rho_+$ was estimated as a finite difference. Updating the cup edge normally moved $a$ away from the discrete $\theta$-lattice points ($n\Delta\theta$). Linear interpolation was then performed to move them back. Finally, for each point on the cup edge, the mean curvature was calculated, giving a value for $\rho_+$, which was then imposed as a boundary condition.

\subsection*{Parameter values}

The parameter values were chosen as follows: $D=0.4\mu\rm{m}^2\rm{s}^{-1}$ is close to the Fc$\gamma$ receptor diffusion constant given in \cite{Howard}; $\rho_0=50\mu\rm{m}^{-2}$ is a typical receptor density \cite{Freund,ReceptorDensity}; $\rho_L=500\mu\rm{m}^{-2}$ is a typical ligand density \cite{LigandDensity}; $\mathcal{E}=15$ is the measured Fc$\gamma$R-IgG binding free energy ($15k_BT$) \cite{BindingValue}; $\mathcal{B}=0.4$--$41$ are typical values of the bending modulus ($0.4k_BT$--$41k_BT$) \cite{BendingModulusValue}. Signalling parameters ($D_S$, $\beta$, $\tau$) are chosen as in our previous paper \cite{OurPaper}.


\section*{Acknowledgments}

DMR and RGE were supported by BBSRC grant BB/I019987/1. DMR was also supported by the Wellcome Trust Institutional Strategic Support Award (WT105618MA). RGE also acknowledges funding from ERC Starting Grant 280492-PPHPI.



\begin{thebibliography}{99}

\bibitem{HourglassII}
Clarke M, Engel U, Giorgione J, M\"uller-Taubenberger A, Prassler J, Veltman D, Gerisch G (2010) Curvature recognition and force generation in phagocytosis. {\it BMC Biol} 8:154.

\bibitem{Borrelia}
Rittig MG, Krause A, H\"aupl T, Schaible UE, Modolell M, Kramer MD, L\"utjen-Drecoll E, Simon MM, Burmester GR (1992) Coiling phagocytosis is the preferential phagocytic mechanism for {\it Borrelia burgdorferi}. {\it Infect Immun} 60:4205--4212.

\bibitem{Legionella}
Horwitz MA (1984) Phagocytosis of the legionnaires' disease bacterium (legionella pneumophila) occurs by a novel mechanism: engulfment within a pseudopod coil. {\it Cell} 36:27--33.

\bibitem{ChampionI}
Champion JA, Mitragotri S (2006) Role of target geometry in phagocytosis. {\it Proc Natl Acad Sci USA} 103(13):4930--4.

\bibitem{ChampionII}
Champion JA, Mitragotri S (2009) Shape induced inhibition of phagocytosis of polymer particles. {\it Pharm Res} 26(1):244--9.

\bibitem{Howard}
van Zon JS, Tzircotis G, Caron E, Howard M (2009) A mechanical bottleneck explains the variation in cup growth during Fc$\gamma$R phagocytosis. {\it Mol Syst Biol} 5:298.

\bibitem{Doherty2009}
Doherty GJ, McMahon HT (2009) Mechanisms of endocytosis. {\it Annu Rev Biochem} 78:857--902.

\bibitem{SwansonReceptorDynamics}
Diakonova M, Bokoch G, Swanson JA (2002) Dynamics of cytoskeletal proteins during Fc$\gamma$ receptor-mediated phagocytosis in macrophages. {\it Mol Biol Cell} 13:402--411.

\bibitem{DartMyosin1G}
Dart AE, Tollis S, Bright MD, Frankel G, Endres RG (2012) The motor protein myosin 1G functions in Fc$\gamma$R-mediated phagocytosis. {\it J Cell Sci} 125:6020--6029.

\bibitem{Smythe1991}
Smythe E, Warren G (1991) The mechanism of receptor-mediated endocytosis. {\it Eur J Biochem} 202:689--699.

\bibitem{GriffinI}
Griffin FM, Griffin JA, Leider JE, Silverstein SC (1975) Studies on the mechanism of phagocytosis: I. Requirements for circumferential attachment of particle-bound ligands to specific receptors on the macrophage plasma membrane. {\it J Exp Med} 142:1263--1282.

\bibitem{GriffinII}
Griffin FM, Griffin JA, Leider JE, Silverstein SC (1976) Studies on the mechanism of phagocytosis: II. The interaction of macrophages with antiimmunoglobulin IgG-coated bone marrow-derived lymphocytes. {\it J Exp Med} 144:788--809.

\bibitem{Samaj2004}
\v{S}amaj J, Balu\v{s}ka F, Voigt B, Schlicht M, Volkmann D, Menzel D (2004) Endocytosis, Actin Cytoskeleton, and Signaling. {\it Plant Physiol} 135:1150--1161.

\bibitem{Qualmann2002}
Qualmann B, Kessels MM (2002) Endocytosis and the cytoskeleton. {\it Int Rev Cytol} 220:93--144.

\bibitem{SwansonSignallingI}
Swanson JA (2008) Shaping cups into phagosomes and macropinosomes. {\it Nat Rev Mol Cell Bio} 9:639--649.

\bibitem{UnderhillOzinsky}
Underhill DM, Ozinsky A (2002) Phagocytosis of microbes: complexity in action. {\it Annu Rev Immunol} 20:825--852.

\bibitem{Sun2005}
Sun SX, Wirtz D (2005) Mechanics of Enveloped Virus Entry into Host Cells. {\it Biophys J} 90:L10--L12.

\bibitem{Tzlil2004}
Tzlil S, Deserno M,Gelbart WM, Ben-Shaul A (2004) A statistical-thermodynamic model of viral budding. {\it Biophys J} 86:2037--2048.

\bibitem{Effenterre2003}
van Effenterre D, Roux D (2003) Adhesion of colloids on a cell surface in competition for mobile receptors. {\it Europhys Lett} 64:543--549.

\bibitem{Zhang2015}
Zhang T, Sknepnek R, Bowick MJ, Schwarz JM (2015) On the modeling of endocytosis in yeast. {\it Biophys J} 108:508--519.

\bibitem{Heinrich1B}
Herant M, Heinrich V, Dembo M (2006) Mechanics of neutrophil phagocytosis: experiments and quantitative models. {\it J Cell Sci} 119:1903--1913.

\bibitem{HeinrichModel2}
Herant M, Lee C-Y, Dembo M, Heinrich V (2011) Protrusive push versus enveloping embrace: computational model of phagocytosis predicts key regulatory role of cytoskeletal membrane anchors. {\it PLoS Comput Biol} 7:e1001068.

\bibitem{DasguptaI}
Dasgupta S, Auth T, Gompper G (2013) Wrapping of ellipsoidal nano-particles by fluid membranes. {\it Soft Matter} 9:5473.

\bibitem{DasguptaII}
Dasgupta S, Auth T, Gompper G (2014) Shape and orientation matter for the cellular uptake of nonspherical particles. {\it Nano Lett} 14(2):687--693.

\bibitem{TollisZipper}
Tollis S, Dart AE, Tzircotis G, Endres RG (2010) The zipper mechanism in phagocytosis: energetic requirements and variability in phagocytic cup shape. {\it BMC Syst Biol} 4:149.

\bibitem{Freund}
Gao H, Shi W, Freund L (2005) Mechanics of receptor-mediated endocytosis. {\it Proc Natl Acad Sci USA} 102(27):9469--9474.

\bibitem{DecuzziFerrari}
Decuzzi P, Ferrari M (2008) The receptor-mediated endocytosis of nonspherical particles. {\it Biophys J} 94(10):3790--3797.

\bibitem{OurPaper}
Richards DM, Endres RG (2014) The mechanism of phagocytosis: two stages of engulfment. {\it Biophys J} 107:1542--1553.

\bibitem{Chithrani2006} Chithrani BD, Ghazani AA, Chan WC (2006) Determining the size and shape dependence of gold nanoparticle uptake into mammalian cells. {\it Nano Lett} 6:662--668.

\bibitem{Hourglass}
Dieckmann R, von Heyden Y, Kistler C, Gopaldass N, Hausherr S, Crawley SW, Schwarz EC, Diensthuber RP, C\^ot\'e GP, Tsiavaliaris G, Soldati T (2010) A myosin IK-Abp1-PakB circuit acts as a switch to regulate phagocytosis efficiency. {\it Mol Biol Cell} 21:1505--18.

\bibitem{Engqvist2003}
Engqvist-Goldstein AE, Drubin DG (2003) Actin assembly and endocytosis: from yeast to mammals. {\it Annu Rev Cell Dev Biol} 19:287--332.

\bibitem{Robertson2009}
Robertson AS, Smythe E, Ayscough KR (2009) Functions of actin in endocytosis. {\it Cell Mol Life Sci} 66:2049--206.

\bibitem{McMahon2015}
McMahon HT, Boucrot E (2015) Membrane curvature at a glance. {\it J Cell Sci} 128:1065--1070.

\bibitem{Hoppe2004}
Hoppe AD, Swanson JA (2004) Cdc42, Rac1, and Rac2 display distinct patterns of activation during phagocytosis. {\it Mol Biol Cell} 15:3509--3519.

\bibitem{Sharma2010}
Sharma G, Valenta DT, Altman Y, Harvey S, Xie H, Mitragotri S, Smith JW (2010) Polymer particle shape independently influences binding and internalization by macrophages. {\it J Control Release} 147(3):408--12.

\bibitem{MycobacteriaEjection}
Hagedorn M, Rohde KH, Russell DG, Soldati T (2009) Infection by tubercular mycobacteria is spread by nonlytic ejection from their amoeba hosts. {\it Science} 323:1729--1733.

\bibitem{DrugShapeI}
Champion JA, Katare YK, Mitragotri S (2007) Particle shape: A new design parameter for micro- and nanoscale drug delivery carriers. {\it J Control Release} 121:3--9.

\bibitem{DrugShapeII}
Nowacek AS, Balkundi S, McMillan J, Roy U, Martinez-Skinner A, Mosley RL, Kanmogne G, Kabanov AV, Bronich T, Gendelman HE (2011) Analyses of nanoformulated antiretroviral drug charge, size, shape and content for uptake, drug release and antiviral activities in human monocyte-derived macrophages. {\it J Control Release} 150(2):204--211.

\bibitem{Huang2010}
Huang X, Teng X, Chen D, Tang F, He J (2010) The effect of the shape of mesoporous silica nanoparticles on cellular uptake and cell function. {\it Biomaterials} 31(3):438--448.

\bibitem{ReceptorDensity}
Quinn O, Griffiths G, Warren G (1984) Density of newly synthesized plasma membrane proteins in intracellular membranes. II. Biochemical studies. {\it J Cell Biol} 98:2142--2147.

\bibitem{LigandDensity}
Gandour DM, Walker WS (1983) Macrophage cell cycling: influence on Fc receptors and antibody-dependent phagocytosis. {\it J Immunol} 130:1008--1012.

\bibitem{BindingValue}
Raychaudhuri G, McCool D, Painter RH (1985) Human IgG1 and its Fc fragment bind with different affinities to the Fc receptors on the U937, HL-60, and ML-1 cell lines. {\it Mol Immunol} 22:1009--1019.

\bibitem{BendingModulusValue}
Pontes B, Ayala Y, Fonseca ACC, Rom\~ao LF, Amaral RF, Salgado LT, Lima FR, Farina M, Viana NB, Moura-Neto V, Nussenzveig HM (2013) Membrane elastic properties and cell function. {\it PLoS One} 8:e67708.

\end{thebibliography}
\end{document}